\documentclass[a4paper,12pt]{article}
\pdfoutput=1
\usepackage{jcappub}
\usepackage{amsthm,graphicx}
\usepackage{epsfig}
\usepackage{latexsym, amssymb} 
\usepackage{amsmath}
\def\beq{\begin{equation}}
\def\eeq{\end{equation}}
\def\br{\begin{eqnarray}}
\def\er{\end{eqnarray}}
\def\benu{\begin{enumerate}}
\def\efnu{\end{enumerate}}

\def\l{\left}
\def\r{\right}

\begin{document}
\title{Witnessing the reionization history using Cosmic Microwave Background observation from Planck}
\author[1]{Dhiraj Kumar Hazra,} 
\author[1,2,3,4]{George F. Smoot}

\affiliation[1]{AstroParticule et Cosmologie (APC)/Paris Centre for Cosmological Physics, Universit\'e
Paris Diderot, CNRS/IN2P3, CEA/lrfu, Observatoire de Paris, Sorbonne Paris Cit\'e, 10, rue Alice Domon et Leonie Duquet, 75205 Paris Cedex 13, France}
\affiliation[2]{Institute for Advanced Study, Hong Kong University of Science and Technology, Clear Water Bay, Kowloon, Hong Kong}
\affiliation[3]{Physics Department and Lawrence Berkeley National Laboratory, University of California, Berkeley, CA 94720, USA}
\affiliation[4]{Energetic Cosmos Laboratory, Nazarbayev University, Astana, Kazakhstan}

\emailAdd{dhiraj.kumar.hazra@apc.univ-paris7.fr, gfsmoot@lbl.gov} 
\abstract 
{We constrain the history of reionization using the data from Planck 2015 Cosmic Microwave Background (CMB) temperature and polarization anisotropy observations. 
We also use prior constraints on the reionization history at redshifts $\sim7-8$ obtained from Lyman-$\alpha$ emission observations.
Using the free electron fractions at different redshifts as free parameters, we construct the complete reionization history using polynomials. 
Our construction provides an extremely flexible framework to search for the history of reionization as a function of redshifts.  
We present a conservative and an optimistic constraint on reionization that are categorized by the flexibilities of the models and datasets used to constrain 
them, and we report that CMB data marginally favors extended reionization histories. 
In both the cases, we find the mean values of optical depth to be larger ($\approx0.09$ and $0.1$) than what we find in standard steplike reionization histories ($0.079\pm0.017$).
At the same time we also find that the maximum free electron fraction 
allowed by the data for redshifts more than 15 is $\sim0.25$ at 95.4\% confidence limit in the case of optimistic constraint. 
}

\maketitle

\section{Introduction}

Photons from bright sources that appear during the onset of structure formation, ionized the intergalactic medium (IGM) gas (mainly hydrogen and helium) which we refer to as the reionization
of the Universe. From the spectra of quasars we know that the Universe is now completely ionized. However, we are yet to make a precise determination of the history of reionization. 
This history is an important factor in determining many evolutionary processes in the development of the modern Universe 
as well as a factor in estimating a number of cosmological parameters such as the amplitude and spectral index of primordial perturbations, neutrino mass. 
We approach to understand the reionization history by using CMB observations -- publicly available Planck 2015 data~\cite{Aghanim:2015xee} -- and by using Lyman-$\alpha$ observations. Using Gunn-Peterson optical depth, dark pixels in quasar spectra, 
Lyman-$\alpha$ damping wings in quasars and Gamma-Ray Bursts (GRB) we understand that the reionization process eventually was fairly complete by 
redshift ($z$) 6~\cite{Bouwens:2015vha}. Due to high rate of absorption by neutral hydrogen clouds, we do not have sufficient detection of quasar spectra at $z>6$ to witness the 
process of reionization. Hence for cosmological parameter estimation using temperature anisotropy data researchers frequently use a symmetric hyperbolic tangent form of 
reionization with a width of 0.5 in redshift. While we do not know the detailed process, a symmetric form is too simple assumption to 
capture a complex process of reionization. At the same time since this form does not allow extended reionization scenarios, our parameter estimation 
becomes biased towards such sharp reionization process. Over time, different methods were proposed to capture the reionization histories. 
Using principal component analysis (PCA)~\cite{Hu:2003gh,Mortonson:2007hq,Mitra:2015yqa}, redshift asymmetric parametrization~\cite{Douspis:2015nca}   
constraints on reionization had been achieved. In this paper, we introduce a new approach where we use the free electron fractions  
at different redshifts as free parameters. We obtain a smooth history of reionization joining the fractions using a piecewise cubic Hermite polynomials.  
We find that this construction allows us to obtain constraints on the reionization histories at different redshifts which is essential to understand
the complete process. The Thomson scattering optical depth is the integrated electron fraction over the line of sight.
Scattering of CMB photons with the free electrons from reionization suppresses the anisotropy power spectrum and modifies 
the large scale CMB polarization power spectrum.
While CMB temperature data allows us to have an overall constraint on the optical depth and the primordial scalar power spectrum amplitude, 
large scale CMB polarization can capture the broad feature of reionization history. 
We provide a conservative bound on the reionization process from CMB by allowing sufficient freedom in our construction. 
With the polarization data, the Planck Collaboration has presented their constraints on reionization in Planck-2015 release~\cite{Ade:2015xua} and Planck-2016~\cite{Aghanim:2016yuo,Adam:2016hgk} 
intermediate release. The intermediate release contains large scale polarization data from Planck High Frequency Instruments (HFI). Since the data and the power spectra 
likelihoods for the 2016 release are not available publicly, in this paper we have only been able to use Planck-2015 likelihoods.
Apart from CMB temperature and polarization data, we use derived constraints on the neutral hydrogen fraction at redshifts $7-8$ from 
Lyman-$\alpha$ emission in galaxies which act as anchors to restrict CMB constraints to provide an optimistic or restricted bound on monotonic reionization history.

This paper is organized as follows. In the next section~\ref{sec:formalism}, we explain our construction of reionization history, discuss the data and priors used in this paper.
We also mention present constraints on the baseline model. In section~\ref{sec:results}, we present the results obtained and provide the samples of reionization histories that 
are presently allowed by the data. Finally, in section~\ref{sec:conclusions}, we discuss future possibilities and conclude.

%%%%%%%%%%%%%%%%%%%%%%%%%%%%%%%%%%%%%%%%%%%%%%%%%%%%%%%%%%%%%%%%%%%%%%%%%%%%%%%
\section{Methodology, datasets and priors}~\label{sec:formalism}
Until now for cosmological parameter estimation using CMB data, usually a simple form of reionization history has been used. Those analyses assume for reionization that the free electron fraction ($x_e(z)$) 
per hydrogen ionization follows a smooth hyperbolic tangent step that connects the electron fractions leftover from the recombination to its value after the complete hydrogen reionization. 
\begin{equation}
x_e(z)=\frac{1+F_{\rm He}}{2}\l[1+\tanh\l(\frac{y(z_{\rm re})-y(z)}{\Delta_{\rm Reion}}\r)\r]~\label{eq:tanhreion},
\end{equation}
where $y(z)=(1+z)^{3/2}$. ${\Delta_{\rm Reion}=1.5\sqrt{1+z_{\rm re}}dz}$ is the width of the step and that quantifies how fast the Universe has changed its state from neutral to ionized. The published 
Planck baseline analysis fixes $dz=0.5$. $1+F_{\rm He}$ in the numerator then includes the electrons from first ionized state of helium alongside
hydrogen reionization. $F_{\rm He}$ is calculated from mass fraction of helium consistent with Big Bang Nucleosynthesis and it approximately takes the value 0.08.
On top of this, there is a second Tanh step representing the second ionization of helium with a redshift width of 0.5 at redshift 3.5 which is added to the free electron fraction~\cite{cambsite,Lewis:1999bs}. 
Signatures of second reionization of helium between $z\sim3-4$ can be found in~\cite{Schaye:1999vr,Theuns:2002vm}. 
$z_{\rm re}$ is the redshift where the free electron fraction from hydrogen becomes exactly half. From this symmetric Tanh reionization model and using 
Planck TTTEEE + lowTEB likelihood we find $z_{\rm re}=10^{+1.7}_{-1.5}$ and the Thomson scattering optical depth is obtained to be $0.079\pm0.017$.

Since our main objective here is to explore to what extent we can constrain the history of reionization as a function of redshift, we do not work with the Tanh form or its variants. 
We introduce a free form history of reionization (poly-reion) utilizing a Piecewise Cubic Hermite Interpolating Polynomials (PCHIP). Our free electron fraction is parametrized by new free 
parameters that act as nodes of the polynomial. $x_e(z)$ is given as: 
\begin{equation}
x_e(z)=(1+F_{\rm He})f(z)~\label{eq:polyreion},
\end{equation}
where $f(z)$ is the polynomial connecting the nodes. In this analysis, we have used 4 free nodes at different redshifts. PCHIP has the property of preserving monotonicity. 
If the electron fractions at the nodes are monotonically increasing with the decrease in redshifts, the polynomials joining the nodes will also respect monotonicity.
However, in this paper, we do not always assume monotonicity in the electron fractions at the nodes and in those cases $x_e(z)$ can show oscillatory behavior. 
Another important advantage with this polynomial is that it does not let the $x_e(z)$ from hydrogen to be more than {\it one} or less than {\it zero} which are physical limits. 
These two factors (monotonicity and physical limits) are in contrast to PCA analysis where one gets oscillations from the eigenfunctions that do not respect these limits.

Since most of the Lyman-$\alpha$ forest measurements~\cite{Bouwens:2015vha,Fan:2005es} indicate that reionization must have completed (neutral hydrogen fraction, $x_{\rm HI}$ goes below~$\sim10^{-4}$) 
latest by $z=5.5$, we impose $f(z=5.5)=1$ in Eq.~\ref{eq:polyreion}. Since the first sources of reionization might not have formed before $z=30$~\cite{firstsources}, we use another
node at $z=30$ where $f(z)$ is fixed to the electron fraction leftover from recombination (approximately $2.5\times10^{-4}$) for a set of background parameters. Keeping the $f(z)$'s at these two nodes fixed, we allow the $f(z)$'s 
at $z=6,~7.5,~10$ and at $20$ to vary as free parameters. We aim to keep the nodes in such times of evolutions around which we have 
observational constraints that we can utilize. Most of the probes around $z\sim6$ using dark pixels in Lyman-$\alpha$ and Lyman-$\beta$ forests~\cite{McGreer:2014qwa},
Gunn-Peterson damping wings~\cite{Schroeder} have reported that cosmic reionization is almost (but not completely) over by $z=6$. 
Likewise GRB observations (limited number of lines of sight) indicate that reionization is atleast halfway complete by $z\sim6.3$~\cite{Totani:2005ng,McQuinn:2007gm}.
Hence we choose one node at $z=6$. 
It has also been demonstrated that evolving visibility of Lyman-$\alpha$ emission in the galaxies between $7<z<8$ detected by Keck MOSFIRE spectrograph can be used to
constrain the reionization history at high redshifts~\cite{Schenker:2014tda}. In order to use the constraints we place another node at redshift 7.5. 
Since measurements at these redshifts only capture a few sightlines, they can not capture the average history of reionization that we use in CMB. Therefore we do not use these constraints throughout. 
The third node is placed at $z=10$ as using the CMB temperature and polarization measurements, Planck obtained $z_{\rm re}\sim10$~\cite{Ade:2015xua}, which marks the midway of reionization.
This value came down to 8.5 when low-$\ell$ polarization from HFI had been used~\cite{Aghanim:2016yuo,Adam:2016hgk}. However, since we have access to Planck 
2015 data, it is likely that a higher value of reionization redshift and optical depth will be supported (independent of the model, as polarization data do not have high signal to noise ratio) and we can expect some constraints on 
$f(z)$ at $z=10$. With HFI Planck CMB data released this may improve. The final node where we allow the $f(z)$ to vary is $z=20$ to capture traces of reionization at high redshifts.

Using our methodology we attempt to constrain the reionization history in a conservative and in an optimistic framework. In the conservative framework, we do not 
respect the monotonic increase of free electron fraction with the decrease in redshift. Such a variation of ionization with redshift is possible if there was an early outburst 
of star formation and AGN activity and this died back before the main star formation activity happened. 
Because of the great energy requirements to keep the IGM ionized at high redshift, some of the IGM could have recombined before being powered to ionization 
by the second wave of star formation. In this framework we use CMB data from Planck. We use CMB TTTEEE and low-TEB data and likelihoods~\cite{Aghanim:2015xee} 
from Planck 2015 data release. Note that the low-EB likelihood correspond to Low Frequency Instrument (LFI) likelihood since the HFI likelihood is not yet publicly available. Apart 
from CMB, we also use a constraint that $f(z)>0.9$ at $z=6$. While we have some constraints from Lyman-$\alpha$ damping wings~\cite{Schroeder} within 
$z\sim6.24-6.42$, in a conservative approach we decided to follow the results in~\cite{Fan:2006dp}. In the optimistic approach, we assume monotonic increase of 
electron fraction with time. Apart from CMB data, here we also use constraints from Lyman-$\alpha$ observations. As we mentioned, the Lyman-$\alpha$ observations
provide the information of the IGM along a single line of sight. Due to high absorption of the photons by the IGM at high redshifts, 
very few Lyman-$\alpha$ emitting sources have been detected until now and hence we have poor statistical record of neutral hydrogen at $z>6$. However, for an 
optimistic analysis, we make use of the constraints on neutral hydrogen from~\cite{Schenker:2014tda} that use models in~\cite{McQuinn:2007dy} to obtain 
the bounds. In the optimistic framework we use these bounds
to provide a stringent constraint on the process of reionization as a function of redshift, as we want to address to what extent we can constrain reionization
using the present day data. In other words, the optimistic bound can be referred as restricted bounds as well.

The prior on the reionization history in the conservative case is imposed only at $z=6$ where we claim $f(z)>0.9$. At the other three nodes the $f(z)$ is allowed to
take values between 0 and 1. In the optimistic case, at $z=6.1$ we impose constraints from Gunn-Peterson optical depth quoted in~\cite{Bouwens:2015vha,Fan:2005es}. 
At $z=7$ we use the $x_{\rm HI}=0.34^{+0.09}_{-0.12}$ and at $z=8$ we impose $x_{\rm HI}>0.65$ from~\cite{Schenker:2014tda}. Other high redshift
probes such as Lyman-$\alpha$ emission in galaxies~\cite{Caruana:2012ww,Ono:2011ew,Pentericci:2014nia,Robertson:2013bq,Tilvi:2014oia},
Lyman-$\alpha$ emitters~\cite{Ota:2007nx}, quasar near zone~\cite{Mortlock:2011va,Bolton:2011vb} also agree with these bounds at $z=7,8$ at 1-2$\sigma$ level. 
We have tabulated the priors used here in Table~\ref{tab:priors}. We should mention that for the optimistic case $x_{\rm HI}(z=6)$ in this table denotes $x_{\rm HI}$
at redshift 6.1, as provided in~\cite{Schenker:2014tda}.
Since we are using monotonicity in the optimistic case we have automatic prior on the upper limit of $f(z=10)$ and $f(z=20)$ 
following the constraints on $x_{\rm HI}(z=8)$. To avoid any sharp transition in the visibility function, we smooth the complete reionization history 
with a $\Delta \ln z=0.02$ (as in~\cite{Mortonson:2009qv}). However, we noted negligible change in the confidence contours of the parameters if the smoothing is not applied. 
This fact ensures that PCHIP provides a smooth history (provided the nodes are not placed too close).

\renewcommand{\arraystretch}{1.1}
\begin{table*}[!htb]
\begin{center}
\vspace{4pt}
\begin{tabular}{| c | c | c |}
\hline\hline
\multicolumn{3}{|c|}{\bf Parameter Constraints}\\
\hline
Priors& Conservative& Optimistic\\
\hline
$z_{\rm begin} (x_{\rm HI}=1)$ 
&30 &30\\
\hline
$z_{\rm end} (x_{\rm HI}=0)$ 
&5.5 &5.5\\
\hline
$x_{\rm HI}(z=6)$
&$<0.1$&$0.99957\pm0.00030$~\cite{Bouwens:2015vha,Fan:2005es}\\
\hline
$x_{\rm HI}(z=7)$
&$-$&$0.34^{+0.09}_{-0.12}$~\cite{Schenker:2014tda}\\
\hline
$x_{\rm HI}(z=8)$
&$-$&$>0.65$~\cite{Schenker:2014tda}\\
\hline\hline
\end{tabular}
\end{center}
\caption{~\label{tab:priors} Priors on the reionization history that has been used in this analysis. $z_{\rm begin}$ and $z_{\rm end}$ denotes the beginning and the end of hydrogen reionization 
and we have provided the hard priors that has been used in this analysis. The other rows denote the fraction of neutral hydrogen used as prior constraints from earlier
analyses.} 
\end{table*}

The free parameters in our analysis are, baryon density $\Omega_{\rm b}h^2$, cold dark matter density $\Omega_{\rm CDM}h^2$, $\theta_{\rm MC}$ (ratio of sound horizon to angular diameter distance), scalar 
primordial spectral amplitude $A_{\rm S}$, spectral tilt $n_{\rm S}$ and four $f(z)$'s representing reionization history. Hereafter, these four $f(z)$'s are denoted as $x_e(z=6)$,
$x_e(z=7.5)$, $x_e(z=10)$ and $x_e(z=20)$ respectively. Note that optical depth $\tau$ is not a free parameter but is derived from the reionization history. We have 
modified CAMB in order to incorporate poly-reion and used CosmoMC\cite{cosmomcsite,Lewis:2002ah} for parameter estimation using Markov Chain Monte Carlo (MCMC). 
Similar to Tanh reionization model, we have added the electrons from second ionization of helium to equation~\ref{eq:polyreion}. The second ionization of helium is modeled using Tanh reionization and the width 
and redshift of reionization used is same as mentioned in Tanh reionization case.
Note that as usual, we allow all the nuisance parameters in the Planck likelihood to vary along with the base background cosmology, power spectrum and reionization
parameters. We fix the sum of neutrino masses to 0.06 eV that is the minimum non-zero number obtained from neutrino oscillation bounds 
for normal hierarchy. This follows Planck baseline analysis.

%%%%%%%%%%%%%%%%%%%%%%%%%%%%%%%%%%%%%%%%%%%%%%%%%%%%%%%%%%%%%%%%%%%%%%%%%%%%%%%
\section{Results and discussions}\label{sec:results}
We start with the obtained bounds on the parameters in poly-reion parametrization. In table~\ref{tab:bounds} we provide the mean and 95.4\% bounds on the parameters. 
In the first column we present the bounds for the conservative case and the optimistic case is listed in the second. $\Omega_{\rm b}h^2$, $\Omega_{\rm CDM}h^2$,
$H_0$ constraints are provided for background parameters. While $\theta_{\rm MC}$ has been used for the MCMC analysis, we provide $H_0$ constraints as they can be 
easily compared with other survey results. Compared to Planck baseline results, we note that these background cosmological parameters remains very similar.
\renewcommand{\arraystretch}{1.1}
\begin{table*}[!htb]
\begin{center}
\vspace{4pt}
\begin{tabular}{| c | c | c | c |}
\hline\hline
\multicolumn{4}{|c|}{\bf Parameter Constraints}\\
\hline
Parameters& Baseline & Conservative& Optimistic\\
\hline
 $\Omega_{\rm b}h^2\times10^{2}$ 
&$2.225^{+0.032}_{-0.030}$
 &$2.226\pm0.030$
 & $2.223\pm0.029$\\
\hline
 $\Omega_{\rm CDM}h^2\times10^{2}$ 
 &$11.98\pm0.29$ 
 &$11.97\pm0.27$
 & $11.99\pm0.28$\\
\hline
 $x_e(z=6)$ 
 &$-$
 &$0.95^{>}_{<}$
 & $0.9995^{>}_{-0.0005}$  \\
\hline
 $x_e(z=7.5)$ 
 &$-$
 &$0.41^{>}_{<}$
 & $0.34^{+0.1}_{-0.12}$  \\
\hline
 $x_e(z=10)$ 
 &$-$
 &$0.24^{+0.33}_{<}$
 & $0.21\pm0.12$  \\
\hline
 $x_e(z=20)$ 
 &$-$
 &$0.16^{+0.13}_{<}$
 & $0.13^{+0.1}_{<}$  \\
 \hline
 $H_0$ 
 &$67.3\pm1.3$
 &$67.2\pm1.2$
 & $67.2\pm1.2$ \\
\hline
$\ln[10^{10} A_{\rm S}]$ 
&$3.094\pm0.066$
&$3.13\pm0.052$
&$3.11\pm0.05$\\
\hline
$n_{\rm S}$ 
&$0.965\pm0.01$
&$0.964\pm0.009$
&$0.963\pm0.009$ \\
\hline
\hline
 $\sigma_8$ 
 &$0.831\pm0.026$
 &$0.837\pm0.022$
 &$0.837\pm0.021$ \\
\hline
 $\tau$ 
 &$0.079\pm0.034$
 &$0.097\pm0.027$
 &$0.087^{+0.026}_{-0.024}$  \\
\hline
 $z_{\rm re}$ 
 &$10^{+3.1}_{-3.2}$
 &$-$
 &$-$  \\

\hline\hline
\end{tabular}
\end{center}
\caption{~\label{tab:bounds}Background, reionization and primordial power spectrum constraints in the case of conservative and optimistic considerations. Note that the error bars represent $95.4\%$
confidence. The `$<$' and `$>$' signs represent the parameter is unbounded from below and above respectively. Note that in both the cases we have constraints on the free electron fraction from hydrogen 
reionization as a function of redshift. The first column represents constraints on Planck baseline model that is publicly available at~\cite{Planck95Table}. Note that the improved constraints 
in the optical depth (and therefore in the spectral amplitude) is due to the use of external constraints and priors in the case of optimistic and conservative cases.} 
\end{table*}
As we mentioned, optical depth $\tau$ is a derived parameter here and we note that the conservative case prefers a higher value of $\tau$ compared to the baseline 
Tanh case (about 1$\sigma$ higher). This result is similar with the optical depth obtained using PCA in~\cite{Heinrich:2016ojb}. Optimistic case however 
%%%%%%%%%%%%%%%%%%%%%%%%%%%%%%%%%%%%%%%%%%%%%%%%%%%%%%%%%%%%%%%%%%%%%%%%%%%%%%%
\begin{figure*}[!htb]
\begin{center} 
\resizebox{240pt}{170pt}{\includegraphics{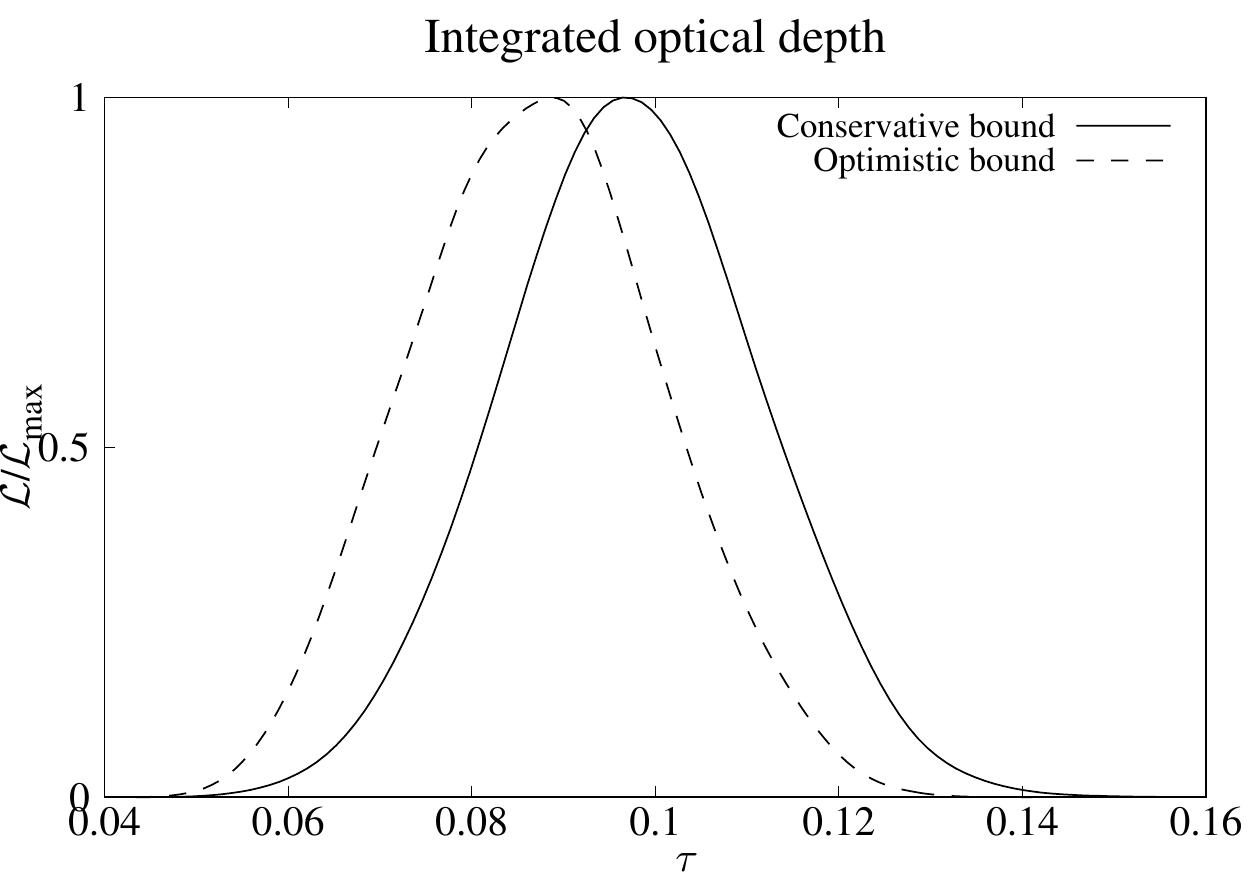}} 
\end{center}
\caption{\footnotesize\label{fig:tau} Marginalized likelihood for the Thomson scattering optical depth. This optical depth is integrated from redshift 0
to the highest redshift where the free electron fraction coincides its leftover value from recombination. We find the conservative case prefers a higher value of
optical depth while the optimistic case prefers a lower value. Note that the optical depth in our analysis is a derived parameter.}
\end{figure*}
prefers a lower optical depth compared to the conservative case, but still more than the baseline case. This is expected as we have some constraints from Lyman-$\alpha$
emission at redshifts 7 and 8 and at the same time we are imposing monotonicity.
In Figure~\ref{fig:tau} we plot the marginalized likelihoods for the optical depth. We note that there is an overall shift in the likelihoods from conservative to the 
optimistic case. Planck HFI large scale polarization favors a much smaller value of optical depth in different parametrizations~\cite{Aghanim:2016yuo,Adam:2016hgk}. While it is expected
that for poly-reion, HFI Planck data may predict a significant lower value of optical depth, it will be more interesting to understand whether such extended 
reionization histories considered here will still be valid. 
\begin{figure*}[!htb]
\begin{center} 
\resizebox{210pt}{150pt}{\includegraphics{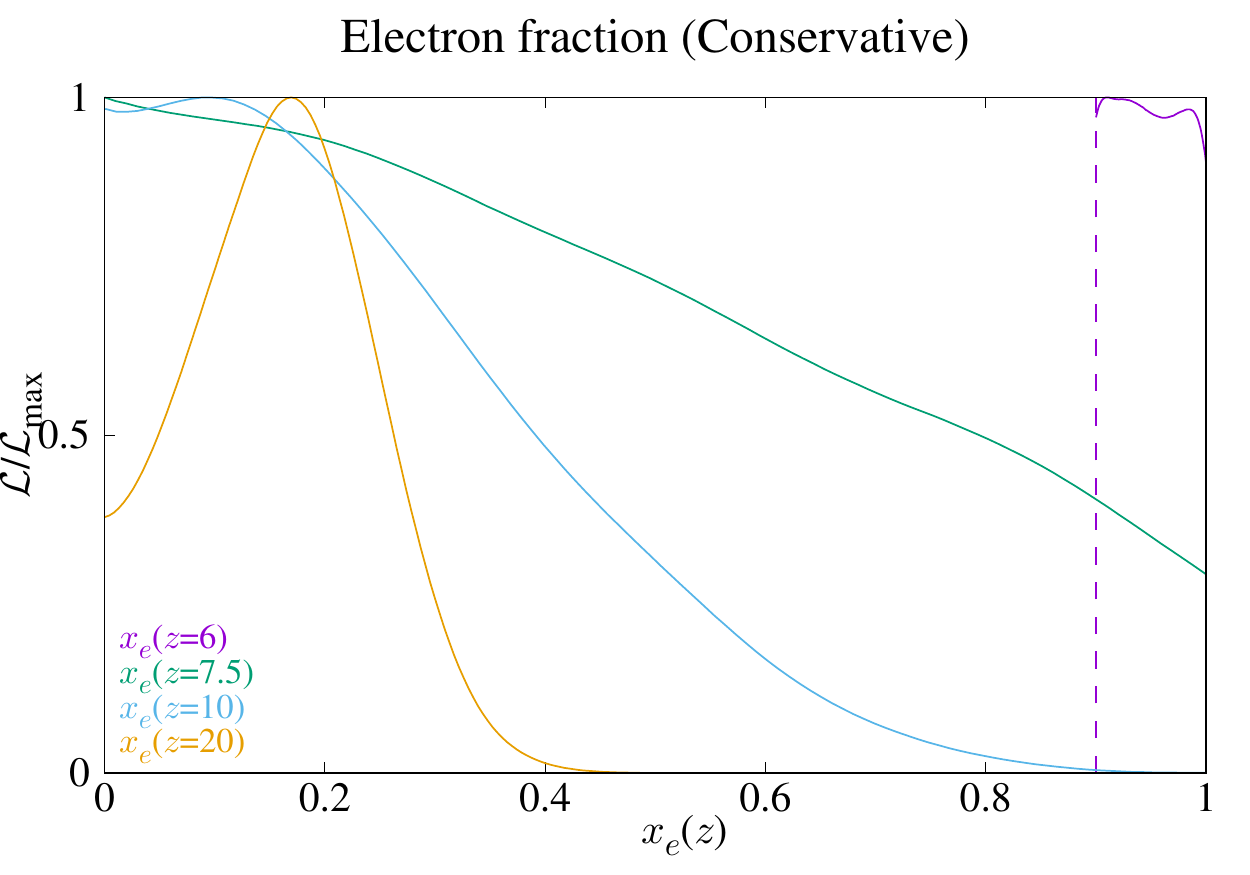}} 
\resizebox{210pt}{150pt}{\includegraphics{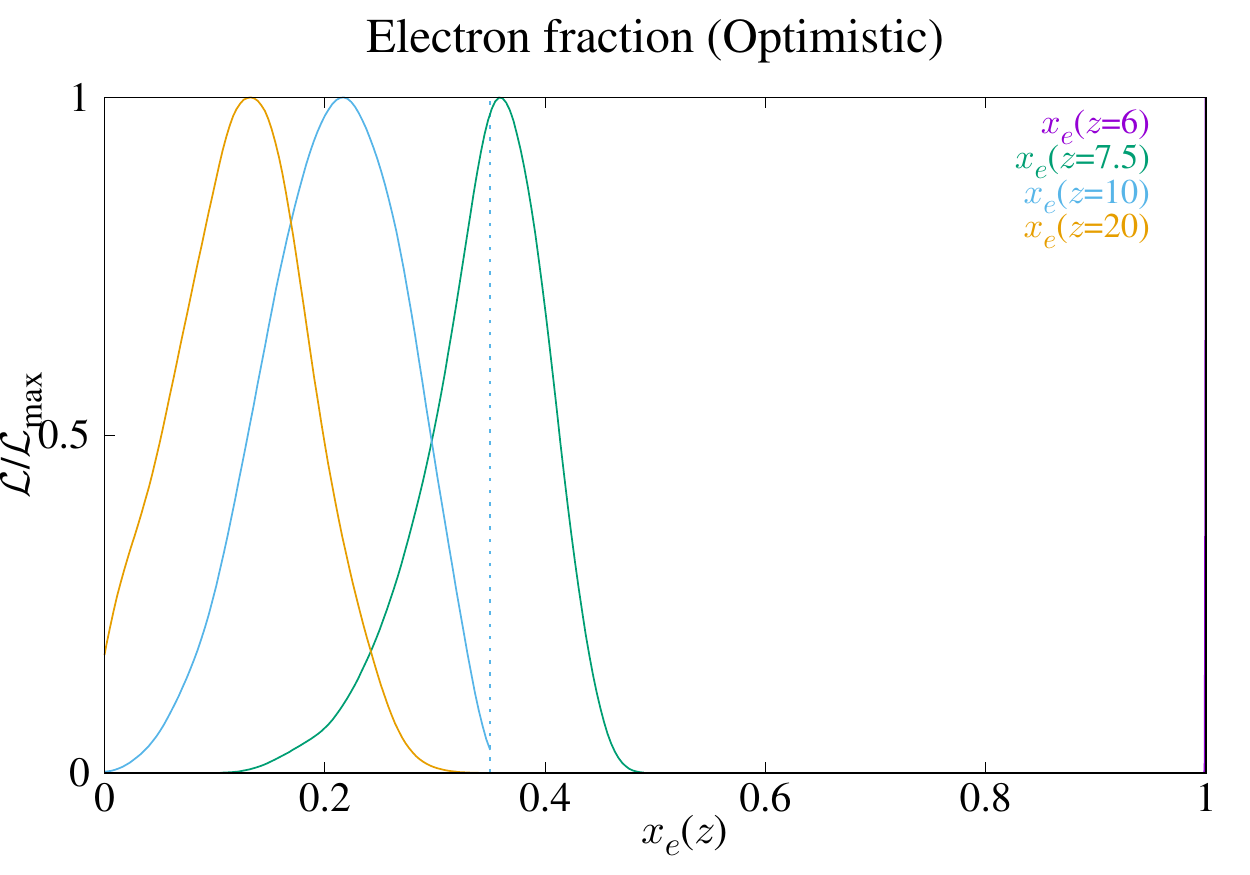}} 

\end{center}
\caption{\footnotesize\label{fig:1Dxe} Marginalized likelihoods for the free parameters of reionization, here the free electron fraction at different redshifts. The left plot shows the 
likelihood for the conservative case and the right one demonstrates the optimistic bounds. The dashed perpendicular lines are priors that were imposed. Note that for optimistic case,
$x_{e}(z=6)$ is within $0.1\%$ of $1$  due to tight constraints from Lyman-$\alpha$ forest observations.}
\end{figure*}
While compared to Tanh reionization we have more parameters in the poly-reion, the bounds on the parameters do not worsen as the extended reionization scenarios are favored 
by the CMB data. There is no noticeable difference in the spectral tilt and the $\sigma_8$ normalization. The change in the spectral amplitude can be attributed to the degeneracies 
with the optical depth and since temperature anisotropies can constrain $A_{\rm S}e^{-2\tau}$, an increased mean value of $\tau$ will support increased mean value of the spectral
amplitude. We have also quoted the reionization redshift $z_{\rm re}$ constraints in the table for baseline model with Tanh reionization. For the poly-reion we do not calculate this 
parameter. Firstly for the conservative case, since the process is not monotonic, the midway of reionization is not restricted to a single redshift. Secondly, for optimistic case,
though the process is monotonic, the $z_{\rm re}$ can not provide an estimate of the reionization process since the $x_e(z)$ is not symmetric {\it w.r.t.} $z_{\rm re}$. Therefore 
different $z_{\rm re}$'s with different duration of reionization can provide the same optical depth.

\begin{figure*}[!htb]
\begin{center} 
\resizebox{200pt}{170pt}{\includegraphics{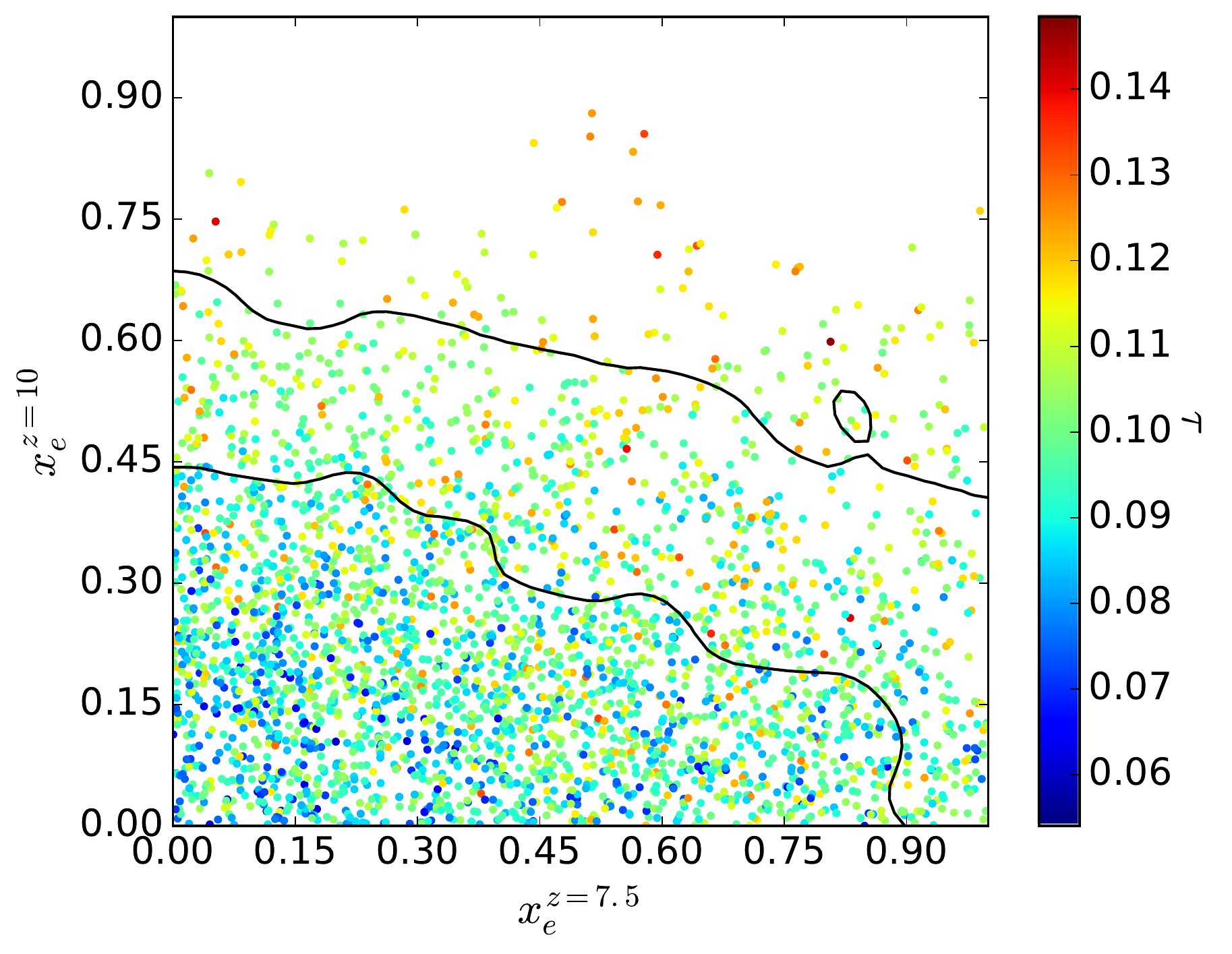}} 
\resizebox{200pt}{170pt}{\includegraphics{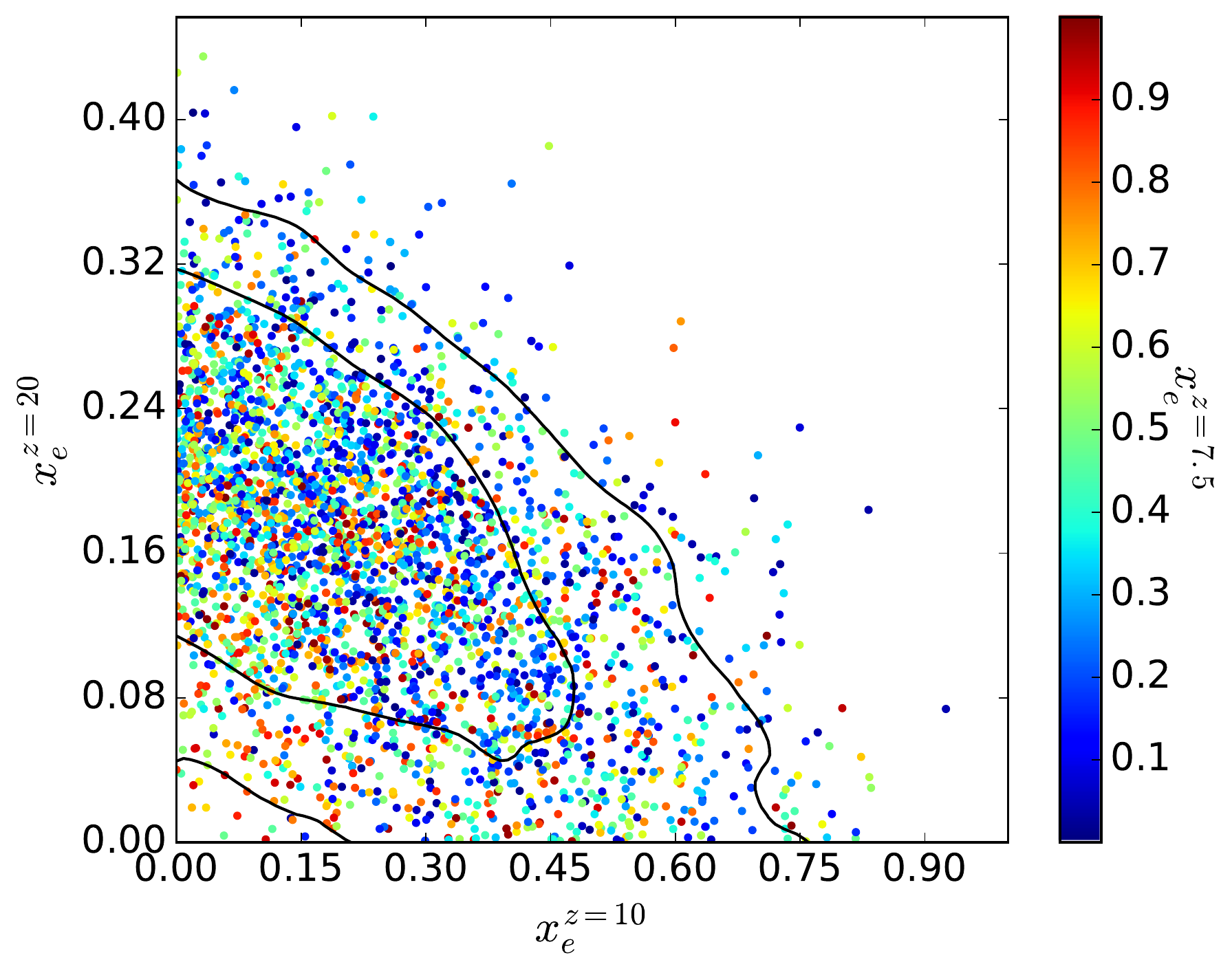}} 

\resizebox{200pt}{170pt}{\includegraphics{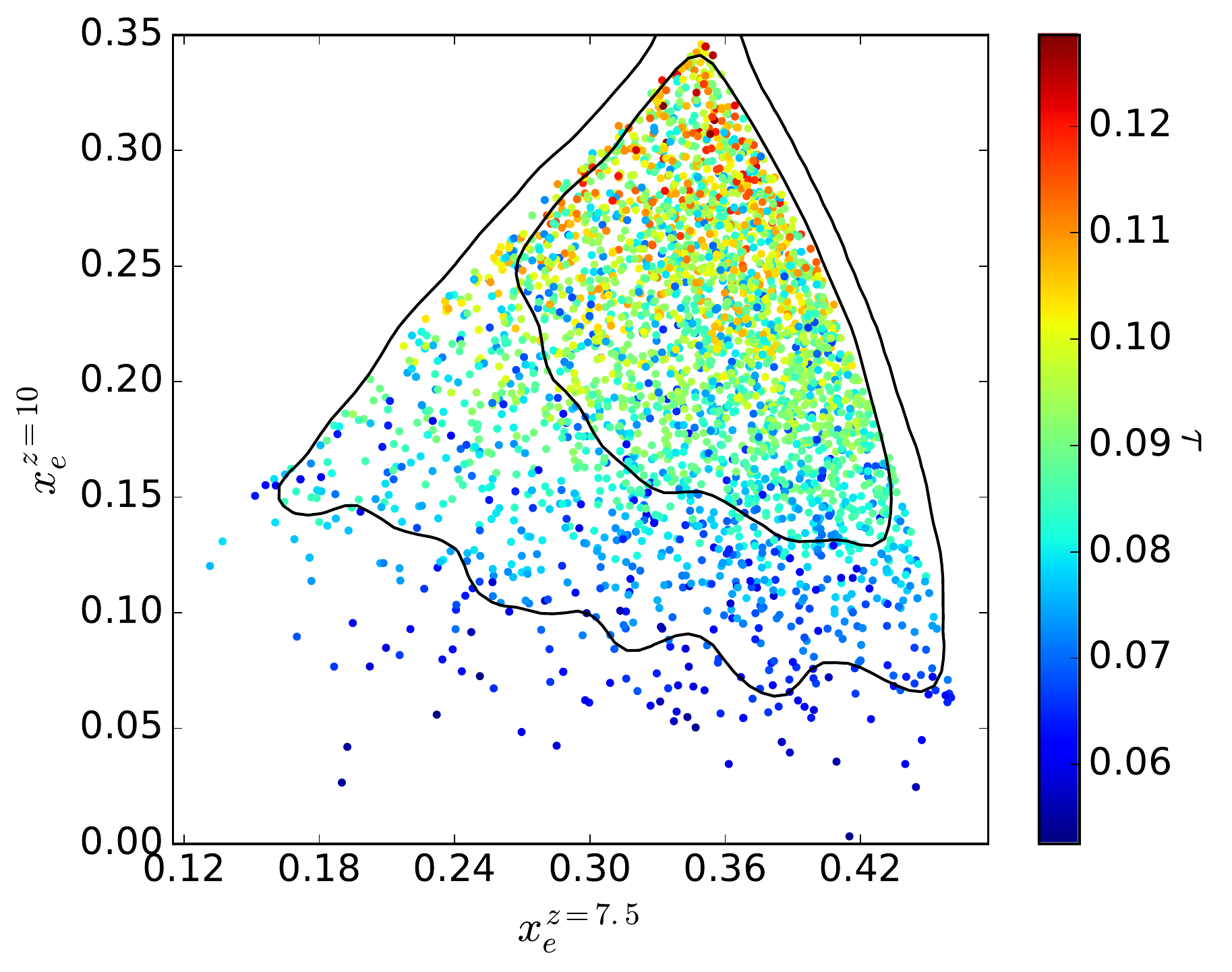}} 
\resizebox{200pt}{170pt}{\includegraphics{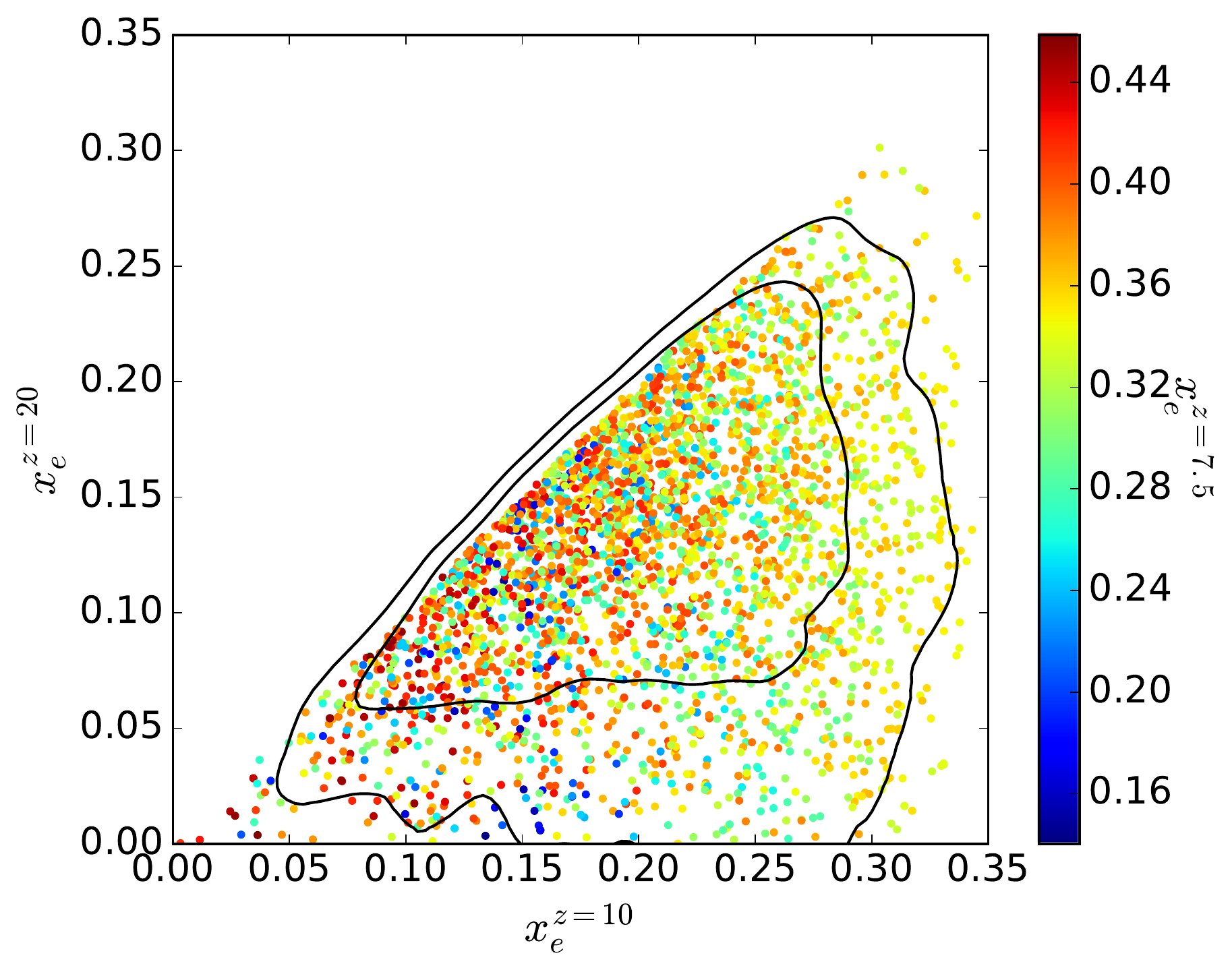}} 
\end{center}
\caption{\footnotesize\label{fig:corr}Bounds on free electron fractions at different redshifts and the Thomson scattering optical depth. Conservative case is plotted in the 
top two plots and the optimistic case is provided below.}
\end{figure*}

Bounds on four parameters describing poly-reion that represent free electron fraction from hydrogen at redshifts 6, 7.5, 10 and 20 are provided in table~\ref{tab:bounds} and the 
corresponding likelihoods are plotted in Figure~\ref{fig:1Dxe}. Since we do not have accurate enough large scale polarization data, it is not expected to have strong bounds in the
conservative case. As we note, at $z=6$ we can not constrain the electron fraction at all. Note that we had imposed a prior of $x_e(z=6)>0.9$. Similarly at redshift 7.5 we do not find 
any constraint at 95.4\% level but the likelihood indicates the data prefers a lower electron fraction which is in favor of incomplete reionization at z=7.5. At redshift 10, 
$x_e(z=10)=1$ is strongly disfavored. Hence with CMB data only, we can rule out end of reionization before $z=10$. As expected, at $z=20$ we do find $x_e(z=20)=1$ is even strongly
disfavored. However there is a mild ($\sim1\sigma$) hint for non-zero electron fraction at $z=20$. Note that this result favors a non-zero optical depth integrated upto high redshift
which is similar to~\cite{Heinrich:2016ojb}. Optimistic case, on the other hand as expected is more constraining. Due to the Gunn-Peterson constraints from~\cite{Bouwens:2015vha,Fan:2005es}
we find strong lower bound on $x_e(z=6)$ and hence it is vanishingly visible in the right plot of this figure. Lyman-$\alpha$ emission constraint at $z=7$ from~\cite{Schenker:2014tda} 
puts tight upper bound on $x_e(z=7.5)$. As we are also using a prior at $z=8$, this prohibits $x_e(z=10)$ and $x_e(z=20)$ to be more than the value at $z=8$ by respecting monotonicity.
We find, with the present data, $x_e(z=10)=0$ is not favored at all in the optimistic case. $x_e(z=20)$ though agrees with an un-reionized state, it prefers a non-zero value 
at a $\sigma$ level that again supports nonzero optical depth at redshifts $>10$ similar to the conservative bounds. 

In order to understand the correlation between the electron fractions at different redshifts and the total optical depth, we plot 2D likelihoods in Figure~\ref{fig:corr}. The top left panel
shows $x_e(z=7.5)$ and $x_e(z=10)$ correlation. The samples are plotted within the 2D contours that are colored by $\tau$. We do not find correlation which indicates in the conservative 
case when we do not have monotonicity as a condition, CMB data can not witness history of reionization between redshift $7.5-10$. At higher redshift however we find strong correlation
in the top right panel. The samples in this plot are colored with $x_e(z=7.5)$ that is not showing much of a correlation with high redshift fractions. Bottom panels show the correlation
between same parameters but in the case of optimistic constraints. Since monotonicity is preserved here and also we have prior constraints on the neutral hydrogen fraction at redshifts 7 and 8, we find
correlation between free electron fractions at $z=7.5,10$ and $20$. The integral $\tau$ is also plotted in colors where the change in $\tau$ with increasing fractions 
is evident. Similarly the bottom right panel that captures the correlations between electron fractions at the three redshifts completely captures the history of reionization. The gradual 
and bounded growths of electron fractions at different redshifts witness the history of reionization. 

\begin{figure*}[!htb]
\begin{center} 
\resizebox{210pt}{150pt}{\includegraphics{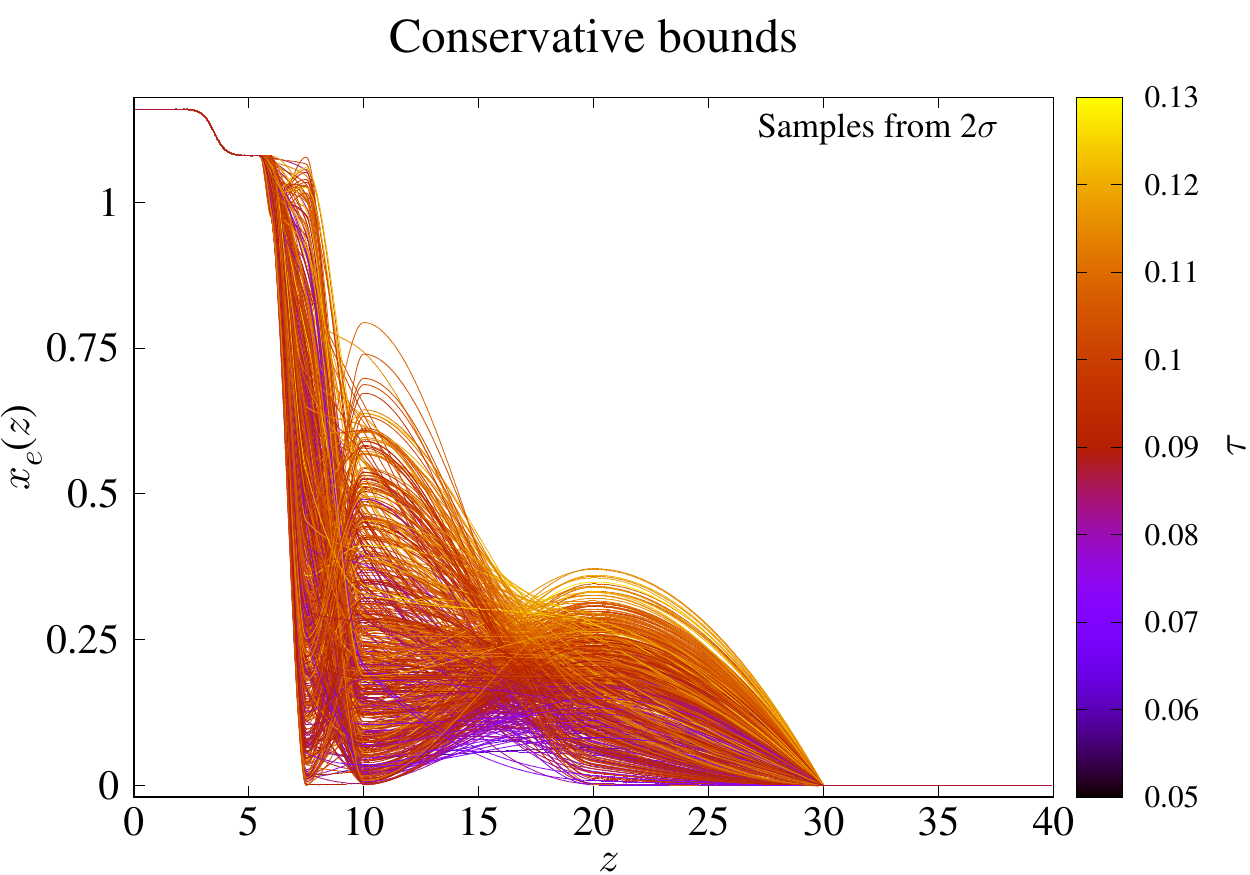}} 
\resizebox{210pt}{150pt}{\includegraphics{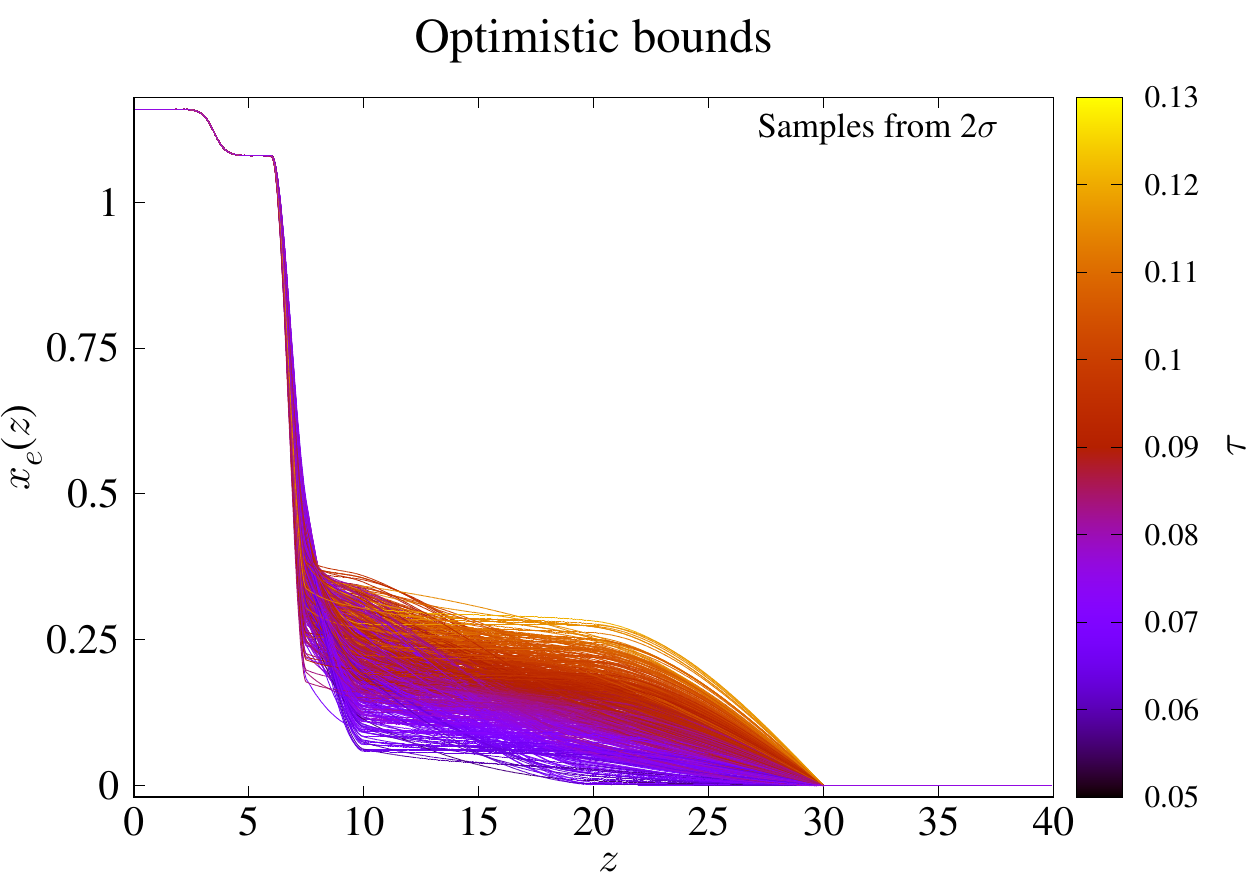}}  
\end{center}
\caption{\footnotesize\label{fig:samples}Samples of free electron fraction within 95.4\% C.L. The left plot shows the samples using the conservative
case where only Planck CMB data has been used and monotonicity in reionization history was not respected while the right plot shows the samples where the Planck CMB and Lyman-$\alpha$ data 
were used respecting monotonic increase of free electron fraction with the decrease in redshift. Note that here electron fractions from the first and the second ionized 
state of Helium are also added that boost up the total free electron fraction to $\sim1.16$ at redshift 0.}
\end{figure*}

Using the samples from the plots in Figure~\ref{fig:corr}, we plot the reionization histories. We provide in Figure~\ref{fig:samples} nearly 200 samples within 95.4\% C.L. in the conservative case (left)
and in the optimistic case (right). In these plot we have added the electron fractions from the both ionization states of helium. Each history is color coded with the optical 
depth integrated till today.
Note that these samples are generated from the PCHIP through the constrained electron fractions at the nodes and hence provide continuous and smooth histories 
from the beginning to the end of reionization. Samples at left do not constrain $x_e(z)$ before $z=10$. As we have noticed in the contours as well, at and after redshift 10
we have upper bounds on the history. Though this plot does not impose monotonicity, we notice decrease in the maximum values on $x_e(z)$ with the increase in redshift.  
A history where reionization has not started at all till $z=10$ is marginally disfavored. The optimistic case evidently is much better constrained. Here too, 
absence of reionization till $z=10$ is not favored. Between $z=10-20$ the samples show a maximum allowed plateau of 25\% reionization. Comparison of two plots 
readily points out that in the optimistic case we can witness the constraints on how the reionization happened. Needless to mention that a revision in the polarization
data from Planck (large scale HFI polarization) will further constrain the history. From the results in~\cite{Adam:2016hgk} one can expect a substantial 
decrease in the mean value of optical depth with values of $\tau>0.1$ being strongly disfavored. Given the colors of the samples, using poly-reion as parametrization,
we expect to bring down the aforementioned plateau to $\sim15\%$ level with HFI data. However, since this projection is only dependent on the bounds on $\tau$,
which in turn was obtained upon assuming Tanh reionization, we must reanalyze poly-reion with upcoming data. 

Poly-reion model provides an improvement of 3-4 in $-2\ln[\rm likelihood]$ compared to standard Tanh reion and thereby marginally favors an extended reionization over
a sharp transition. This improvement comes both from large scale and small scale CMB data. Extended reionization increases the large scale polarization angular 
power spectrum. Observed EE angular power spectra from Planck 2015 release show larger power at large scales ($\ell\sim10-40$) compared to the baseline best fit. 
Poly-reion is able to provide a better match by increasing the large scale power. Having said that, we must admit that at the current state, the data may contain large 
noise and systematic effects and hence without a clean polarization data, it is not possible to have any statistically significant result.

\section{Conclusions}\label{sec:conclusions}
In this paper we discuss a model of extended reionization of Universe. We use piecewise cubic Hermite polynomials to interpolate between the free electron fractions at
different redshifts acting as free parameters. The redshift nodes are chosen in a way to incorporate wide range of histories that can be potentially constrained with the Lyman-$\alpha$ data at high redshifts
and CMB. Our parametrization allows diverse possibilities of reionization by allowing the free electron fraction from hydrogen within the physical range 
and at the same time by respecting monotonicity if demanded. We present two analyses, namely conservative and optimistic. In the conservative reionization histories we do not
respect monotonicity and only constrain them with CMB data while the optimistic histories are constrained with CMB and different Lyman-$\alpha$ data at redshifts higher 
than 6 and assuming that electron fraction from reionized hydrogen is increasing monotonically with time. We find that in both the cases the data lean towards a higher 
value of optical depth with the conservative case suggesting a value of mean optical depth. For the conservative case we find $\tau=0.097\pm0.014$ and for the optimistic case 
we find $\tau=0.087^{+0.013}_{-0.014}$ at 68.3\% C.L. At the same C.L., Planck-2015 constraint for the baseline model with Tanh reionization is $\tau=0.079\pm0.017$~\cite{Ade:2015xua}
while the updated Planck constraints using large scale polarization from HFI is $0.0596\pm0.0089$~\cite{Adam:2016hgk}. The prior constraints imposed in this analysis 
helps to improve the constraint on $\tau$ compared to the Planck 2015 baseline results. We find while only with CMB it is not possible to tightly constrain the
reionization histories, the data marginally favors an extended history. The optimistic case is more constraining
as we use information from Lyman-$\alpha$ emission in galaxies at high redshifts ($z\sim7-8$). There along with monotonicity, we witness moderate to tight constraints on 
the electron fractions with the decrease in redshift. In this case we find, at 95.4\% C.L., maximum allowed electron fraction at $z>15$ is $0.25$ and at the same time 
a {\it zero} value is marginally disfavored. There are two factors that is likely to change our results. Firstly, due to the systematic effects and noise in the LFI large scale 
polarization data, it is possible to have their artefacts within the constraints. The new HFI data favors a lower value of optical depth. With the new data release, using our 
model along with constraining optical depth to a better accuracy, we expect to have much better constraints on high redshift reionizations. Secondly, in the optimistic 
case as we do not have Lyman-$\alpha$ observations for sufficient lines of sight covering large parts of sky, future detections have large scope to improve the constraints.  
Finally, we would like to point out that the parametrization introduced in this paper allows smooth and significantly flexible histories of reionization that capture 
fine changes in this process. At the same time two constraints discussed here clearly show the potential of this model to trace the true history of reionization with the 
upcoming data.

%%%%%%%%%%%%%%%%%%%%%%%%%%%%%%%%%%%%%%%%%%%%%%%%%%%%%%%%%%%%%%%%%%%%%%%%%%%%%%%

\section*{Acknowledgments}
DKH and GFS acknowledge Laboratoire APC-PCCP, Universit\'e Paris Diderot and Sorbonne Paris Cit\'e (DXCACHEXGS) and also the financial support of the 
UnivEarthS Labex program at Sorbonne Paris Cit\'e (ANR-10-LABX-0023 and ANR-11-IDEX-0005-02). The authors would like to thank Eric Linder for his comments on the manuscript. 
DKH would like to thank Suman Majumdar, Sourav Mitra for helpful discussions.

%%%%%%%%%%%%%%%%%%%%%%%%%%%%%%%%%%%%%%%%%%%%%%%%%%%%%%%%%%%%%%%%%%%%%%%%%%%%%%%

%%%%%%%%%%%%%%%%%%%%%%%%%%%%%%%%%%%%%%%%%%%%%%%%%%%%%%%%%%%%%%%%%%%%%%%%%%%%%%%
\end{document}